\documentclass[runningheads]{llncs}
\usepackage{amssymb}
\usepackage{graphicx}
\usepackage{xcolor}
\usepackage{multicol}
\usepackage{multirow}
\usepackage{makecell}
\usepackage{tabularx}
\usepackage[bottom]{footmisc}
\begin{document}
\title{Future Vision of Dynamic Certification Schemes for Autonomous Systems}
%
%
\author{Dasa Kusnirakova \and Barbora Buhnova}


%
%
\institute{Faculty of Informatics, Masaryk University\\ Brno, Czech Republic\\
\email{\{kusnirakova, buhnova\}@mail.muni.cz}}
\maketitle              
\begin{abstract}
As software becomes increasingly pervasive in critical domains like autonomous driving, new challenges arise, necessitating rethinking of system engineering approaches. The gradual takeover of all critical driving functions by autonomous driving adds to the complexity of certifying these systems. Namely, certification procedures do not fully keep pace with the dynamism and unpredictability of future autonomous systems, and they may not fully guarantee compliance with the requirements imposed on these systems. 

In this paper, we have identified several issues with the current certification strategies that could pose serious safety risks. As an example, we highlight the inadequate reflection of software changes in constantly evolving systems and the lack of support for systems' cooperation necessary for managing coordinated movements. Other shortcomings include the narrow focus of awarded certification, neglecting aspects such as the ethical behavior of autonomous software systems. The contribution of this paper is threefold. First, we analyze the existing international standards used in certification processes in relation to the requirements derived from dynamic software ecosystems and autonomous systems themselves, and identify their shortcomings. Second, we outline six suggestions for rethinking certification to foster comprehensive solutions to the identified problems. Third, a conceptual Multi-Layer Trust Governance Framework is introduced to establish a robust governance structure for autonomous ecosystems and associated processes, including envisioned future certification schemes. The framework comprises three layers, which together support safe and ethical operation of autonomous systems.

\end{abstract}
\section{Introduction}
According to the latest estimates, autonomous vehicle (AV) technology would replace the majority of human driving by the year 2050~\cite{litman2022}. Aside from the benefits such as increased road safety, significant cost savings, and lower energy consumption and pollution~\cite{dia2020}, the introduction of self-driving vehicles into public places introduces new challenges for safeguarding the mobility ecosystem as a whole in order to ensure their safe operation. As a result, with rapidly evolving autonomous software improvements, it is necessary to establish regulatory frameworks that are designed to adapt to technological changes in order to reduce safety hazards.

Regulations must unavoidably develop to keep up with technological innovation, especially given the rising software complexity of self-driving systems. This has happened in the past with certification processes. When vehicles were built entirely of mechanical components, such as mechanical brakes or steering, the driving function and all decision-making were in the hands of the driver. In this circumstance, traditional certification procedures were enough. 

However, with the introduction of ABS\footnote{Anti-lock Braking System; A safety system used on land vehicles that is activated in the event of a skid to allow the driver to maintain more control over the vehicle.} and other systems with a higher level of complexity, it became clear that the traditional approach was insufficient in assessing all safety-relevant aspects due to the large number of potential testing scenarios. As a result, process- and functional-oriented safety audits were implemented, one of which being Annex 6 of UN Regulation No.79~\cite{un-reg-79}.

Future road cars are predicted to increasingly replace more vital driving duties, eventually replacing manual driving entirely. This shifts the driver's emphasis and duty to the technology placed in the AV. As a result, the relevance and complexity of electronic control systems used in automobiles will continue to grow. Such a change significantly expands the number of driving scenario possibilities. However, the traditional testing step for manually driven vehicles, that is, confirming the system based on a preset set of tests, will only be able to properly evaluate a small portion of all safety cases and situations~\cite{kalra2016}.

Furthermore, autonomous systems, i.e. \textit{"systems changing their behaviour in response to unanticipated events during operation"}~\cite{watson2005autonomous}, to which AVs clearly belong, rely on automatic software updates at run-time to adapt to specific environment or context changes~\cite{deco2021}, or to improve AI components of an AV in general~\cite{cert-eu-report2020}. Again, typical certification processes seem not to be geared to deal with this problem quickly. Supposing they assume no, or only a small number of modifications in existing approved systems. However, just as software engineering does not end with system deployment, future AVs will require unique technological and legal techniques for full quality control even during run-time to maintain public road safety.

The difficulties of AV certification extend beyond the safety of autonomous ecosystems. According to studies, trust is critical in the adoption of autonomous systems~\cite{trust-cioroaica2019}, from both a societal (people are willing to accept and use the systems) and a technological (interactions between communicating systems during run-time are trustworthy) standpoint~\cite{survey-trust-mng2022}. However, the continued technological advancement of AV beyond the boundaries of previously defined safety requirements impedes trust formation in these systems~\cite{trust-cioroaica2020}, and certification, which is widely perceived as a trust-building mechanism, currently fails to provide sufficient legal guarantees. As a result, it is vital to adapt certification processes to anticipated technical improvements and to promote the integration of AV into society.

In this work, we first investigate the need to enhance current certification processes for autonomous driving systems, and then we compare previously established standards to the requirements placed on future autonomous systems. We describe a model for improving the certification methods, taking into account the experience from social computing as well as the features of dynamic autonomous ecosystems in which such systems operate. To this end, we outline the needs for future autonomous system certification procedures, which need to take a new spin on building trust in dynamic autonomous systems, then we assess current certification standards against the specified requirements and finally present our future vision of dynamic certification schemes through a Multi-Layered Trust Governance Framework to support dynamic certification.

The rest of the paper is organized as follows. Chapter 2 surveys the related work in terms of improving certification standards. Chapter 3 introduces the certification context. Chapter 4 describes the selected standards, the evaluation results of their comparison against the defined certification contexts are presented in Chapter 5. Chapter 6 presents the key suggestions for improvements. The Multi-Layer Trust Governance Framework to Support
Dynamic Certification is introduced in Chapter 7.

\section{Related Work}

There have already been some concerns in the literature on whether the current vehicle certification processes are suitable for the future. However, there is currently no paper that specifically highlights the requirements for certification of dynamic and largely changing autonomous systems, spanning from technical concerns to aspects such as trust and ethics.

Bonnin~\cite{bonnin2018} has identified the need for changes in the certification process, but their critique primarily focuses on the standards' inadequate incorporation of technological advancements in connected software development processes. However, this article does not delve into the specific characteristics of autonomous vehicles and the ecosystems in which they operate. Other criticisms of certification procedures have also been identified. In~\cite{burzio2018}, for instance, the authors call for modifications from a cybersecurity standpoint. However, the most commonly addressed certification problems are those related to the safety of autonomous vehicles, as discussed in~\cite{zhao2022} and~\cite{cummings2019}. The latter paper specifically criticizes the potential decrease in vehicle safety unless software upgrades are considered in the certification process.

The identified deficiencies and criticisms of the current certification procedures have prompted the development of suggestions for improvements. UNECE's GRVA group, in collaboration with experts from OICA, introduced a new approach for validating autonomous vehicles in the context of certification, which involves a multi-pillar approach with a scenario catalog~\cite{unece2022}. Dynamic certification based on modeling and testing, intertwined throughout the system's life cycle, is proposed by Bakirtzis et al.~\cite{dynamic2022}. Paper~\cite{fisher2018} discusses a verifiably-correct dynamic self-certification framework for autonomous systems, while~\cite{hussein2021} introduces a certification framework for autonomous driving systems based on the Turing test. García-Magariño et al.~\cite{garcia2019} propose the use of digital certificates along with trust and reputation policies to ensure safety and detect hijacking vehicles.

Although these attempts partially address the shortcomings of static certification, the debate primarily revolves around safety and security, neglecting other aspects such as ethics and trust in autonomous vehicles. While trust and ethics are often discussed in relation to privacy preservation~\cite{lai2021} and general calls for adjusting software development standards and best practices with social responsibility~\cite{kwan2021,myklebust2020}, their direct consideration within the certification process remains unexplored.

If we are to trust systems that make crucial decisions for us not only in terms of safety but also in ethics, it is imperative to ensure that these aspects are taken into account during the vehicle certification stage, providing legal guarantees.

\section{Specification of the Certification Context}
\label{sec:specs}
Achieving responsible certification of systems becomes challenging when done in isolation from the system's operational environment and without considering the specific characteristics of the system itself.

Autonomous vehicles (AVs) are classified as cyber-physical systems (CPSs), which are defined as \textit{”smart networked systems with embedded sensors, processors and actuators that are designed to sense and interact with the physical world”}~\cite{cps-def}. Consequently, understanding the characteristics of CPSs can assist in identifying the specific traits of AVs that are essential for comprehensive certification.

The objective of this section is to explore the fundamental attributes of autonomous CPSs and the encompassing ecosystems. By establishing the certification context based on this understanding, we can then outline the requirements for certification standards applicable to future autonomous systems.

\subsection{Autonomous cyber-physical systems}
When establishing requirements for certification frameworks for driving systems, it is crucial to take into account the specific characteristics of autonomous systems.

Weyns et al.~\cite{weyns2021} outline key principles for future cyber-physical system (CPS) engineering, which can provide insights into the characteristics of future autonomous systems. The following principles are highlighted:
(1)~\textit{crossing boundaries} refers to the close interaction between social, physical, and cyber spaces;
(2)~\textit{leveraging the humans} emphasizes the integration of humans in the design and operation processes, rather than treating them solely as system users;
(3)~\textit{on-the-fly coalitions} involve forming multi-agent systems to address complex problems;
(4)~\textit{dynamically assured resilience} focuses on the ability to withstand uncertainty, contextual changes, or disruptions and continue providing services;
(5)~\textit{learn novel tasks} emphasizes the effective utilization of past knowledge to handle new and unfamiliar situations.

\subsection{Dynamic software ecosystems}
Furthermore, it is essential to consider the ecosystem in which the autonomous system operates. The ecosystem encompasses the surroundings and entities that interact with the system, defines their relationships, and establishes the system's context, including its primary purpose and usage.

In~\cite{deco2021}, the authors introduce the concept of ecosystems formed by software autonomous systems, which support dynamic, intelligent, and autonomous functionalities required by modern software systems. When designing certification frameworks suitable for autonomous systems, the following features need to be taken into account:
(1) \textit{automation} refers to the automated self-adaptation of the ecosystem in response to contextual changes during runtime;
(2) \textit{autonomy} involves intelligent adaptation to dynamic requirements, aiming to achieve proposed and explicitly defined goals; 
(3) \textit{dynamic goal evaluation} as intelligent adaptation to dynamic needs  with the intention to achieve the proposed and explicitly defined goals; 
(4) \textit{automated trust management} is a critical concept driving decision-making within the ecosystem, as trust guarantees are necessary for collaboration and the fulfilment of goals in dynamic and unpredictable environments; and 
(5) \textit{architecture implications} arising due to the ecosystem's dynamic nature, driven by the continuous connection and disconnection of nodes.

\subsection{Ethical aspects}
When it comes to human-operated vehicles, drivers are responsible for adhering to basic safety and ethical rules while driving. However, the integration of autonomous systems into ecosystems shared with humans changes the ethical dynamics of driving. In such cases, when humans are no longer in control, the autonomous system itself bears a certain moral responsibility.

One well-known ethical dilemma is the trolley problem~\cite{trolley}, where a collision is inevitable, and any action taken results in tragedy. Other examples of ethical considerations arise in situations where altruistic behaviour is applied, such as informing other road users about unseen dangers (e.g., a person crossing the street on a red light). Additionally, solidarity with surrounding entities can be demonstrated by transparently notifying other drivers of changes in speed or direction, promoting smooth traffic flow and avoiding dangerous situations.

Therefore, in addition to the primary requirements of ensuring safety and security~\cite{safety-and-security-first}, ethical considerations play a crucial role in the grey area where some level of harm may still occur, or actions can be taken to benefit the overall ecosystem, even if they are not legally required.

The concept of integrating ethical principles into autonomous driving goes beyond the requirement of \textit{leveraging the humans} in cyber-physical systems (CPS). When ethical standards are applied, individuals are no longer seen merely as users but as equal members of the ecosystem, with their own unique needs and dignity.

Research widely acknowledges the significance of applying ethical norms for establishing trust in autonomous systems~\cite{towards-kwan-2021,ethics-wang-2022,virtue-gerdes-2020}. Given the ethical diversity observed in human society, it is crucial to allow for the adjustment of a system's moral settings to align with individual preferences within a personal ethical framework. However, an unregulated personal ethical framework would lead to numerous ethical dilemmas. Therefore, a preferred ethical model should reflect widely accepted rational ethical inclinations while providing users with limited freedom to customize the ethical settings according to their personal preferences~\cite{ethics-wang-2022}.

This paper does not delve into the specifics of implementing particular ethical principles in autonomous systems. Instead, its focus is on examining whether current certification standards for autonomous vehicles adequately consider ethical issues related to the driving task. The paper aims to contribute to the discussion on the readiness of existing certification standards rather than determining the form and manner of implementing specific ethical principles in autonomous systems.

\section{Evaluation Process}
In order to identify the gaps in the certification processes for autonomous systems, first the requirements on these systems must be formulated, together with a set of standards for evaluation. We determine the criteria based on both the traits of autonomous systems and the traits of the environments in which they operate. We obtain this information from the literature analyzing autonomous systems in dynamic software ecosystems. In terms of choosing certification standards, we chose standards that have been published or re-confirmed within the past seven years (2016-2022) in a variety of automotive-related sectors from international organizations and assessed their suitability for fully autonomous driving.

\subsection{Requirement selection}
The systems themselves and the dynamic software ecosystems in which autonomous systems operate, both of which are described in Section~\ref{sec:specs}, necessitate a change from staticity to a more dynamic approach. Static certification requirements are those that no longer apply in terms of:

\begin{enumerate}
    \item \textit{time}, that a certificate for a vehicle is given based on its state at a specific time point, usually when it is produced,
    \item \textit{context}, that a certificate is issued based on a set of predetermined and limited number of tests,
    \item \textit{collaboration}, that the standard neither supports dynamic creation of coalitions nor considers communication with other entities in the environment,
    \item static \textit{tools}, meaning only documentation review, model report evaluation, or visual inspection, are employed during the certification process.
\end{enumerate}

We propose the transformation of the aforementioned static qualities into their dynamic counterpart, towards which we analyze the chosen standards to examine their suitability for autonomous systems. By this, we believe future certification standards should be able to better reflect the dynamicity of ecosystems. We particularly discuss the dynamicity in the following areas:

\begin{enumerate}
    \item \textit{time}, the standard can handle dynamic system changes, such as those brought on by a software update,
    \item \textit{context}, the standard advises employing tools to completely test the systems' functioning in dynamic context, such as unpredictable and unforeseen circumstances,
    \item \textit{collaboration}, the standard facilitates the dynamic formation of multi-agent systems to enable optimization and more effectively solve complex issues, such as through on-the-fly coalitions,
    \item \textit{tools}, to determine if the tools used to confirm the systems' compliance with the standard are dynamic.
\end{enumerate}

Along with the dynamic features, for the certification standards analysis we also consider

\begin{enumerate}
  \setcounter{enumi}{4}
  \item \textit{ethics}, meaning whether ethical matters are explicitly discussed or taken into consideration within the cases addressed in the particular vehicle certification standard under evaluation.
\end{enumerate}

\begin{figure*}[t]
\centering
\includegraphics[width=330px]{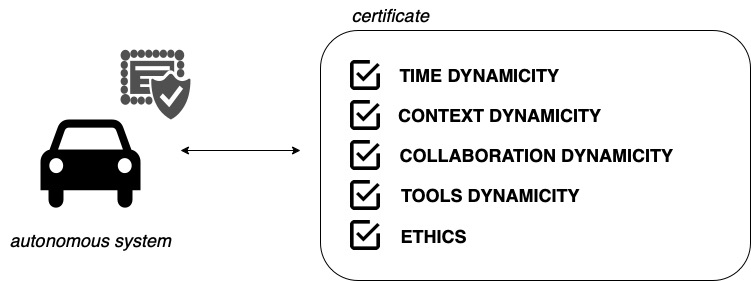}
\caption{Key certification standards' aspects on future autonomous systems~\cite{kusnirakova2023ENASE}}
\label{fig:reqs}
\end{figure*}

Figure~\ref{fig:reqs} illustrates the proposed essential certification standards characteristics for future autonomous systems.

\subsection{Standards selection}
\label{sec:standard-selection}
The following criteria were used to select the documents for review. In the beginning, we looked for standards that had recently been released or reaffirmed by international organizations for standardization (ASAM, ISO, ITU, SAE, UNECE). We concentrated on standards that have been released for light-duty vehicles (passenger cars) in the area of on-road automated driving, and related to the definition of software requirements, software upgrades, testing, validation, and verification, or general guidelines for automated driving. 

During this searching process, we found 27 documents that were relevant. Not all of the selected documents, however, could be examined. We excluded documents that only defined the taxonomy and terminology, exchange format or language specifications (ASAM standards), as well as documents that were not specifically related to automated driving systems (such as audit and other control activities guidelines, such as those in ISO/PAS 5112), which were not publicly accessible at the time this paper was being written. Table~\ref{table:description} contains the final list of standards chosen for review.

\section{Evaluation Results}
Using the criteria outlined in Section~\ref{sec:standard-selection}, we evaluated the standards' appropriateness for fully autonomous driving after choosing the relevant documents. 
The evaluation results are shown in Table~\ref{table:evaluation} with the standards listed alphabetically by their identifiers. The assessment summary is provided at the conclusion of this section. The evaluation results for each of the examined elements are explained in greater detail in the following paragraphs.


\begin{table}[!ht]
\footnotesize
\centering
\begin{tabular*}{\textwidth}{c l l l}
\hline
\textbf{\hspace{1pt}No.\hspace{3pt}} & \textbf{Standard ID \hspace{29pt}} & \textbf{Name \hspace{118pt}} & \textbf{Field}    \\ 
\hline
\hline

1            & ISO 11270         & \makecell[l]{Intelligent transport systems - \\Lane keeping assistance systems \\(LKAS) — Performance require-\\ments and test procedures} & Safety\&Assurance \\ \hline
2            &    ISO 15037 series    &     \makecell[l]{Road vehicles — \\Vehicle dynamics test methods}                                                                                                                  &   Safety\&Assurance                \\ \hline
3             &       ISO 21448    &      \makecell[l]{Road vehicles — \\Safety of the intended functionality}                                                                                                                 &    Safety\&Assurance               \\ \hline
 4            &      ISO 22735            &    \makecell[l]{Road vehicles — Test method to \\evaluate the performance of \\lane-keeping assistance systems}                                                                                                                   &          Safety\&Assurance         \\ \hline
  5           &      ISO 22737             &      \makecell[l]{Intelligent transport systems — \\Low-speed automated driving \\systems for predefined routes — \\Performance requirements, system \\requirements and performance \\test procedures}                                                                                                                 &       Safety\&Assurance            \\ \hline
 6            &      ISO/DIS 26262 series             &  \makecell[l]{Road vehicles — Functional safety}                                                                                                                     &     Safety\&Assurance              \\ \hline
  7           &      ISO/SAE 21434             &         \makecell[l]{Road vehicles — \\Cybersecurity engineering}                                                                                                              &                  Cybersecurity \\ \hline
  8           &    ISO/TR 21959 series               &        \makecell[l]{Road vehicles — Human perfor-\\mance and state in the context of \\automated driving}  &   Human factor     \\ \hline
  9           &  ISO/TR 4804      &     \makecell[l]{Road vehicles — Safety and cyber-\\security for automated driving \\systems — Design, verification \\and validation}    &    \makecell[l]{Safety\&Assurance, \\Cybersecurity}   \\ \hline
  10           &   ISO/TS 5255 series     &     \makecell[l]{Intelligent transport systems — \\Low-speed automated driving \\system (LSADS) service}    &   Data    \\ \hline
  11           &   SAE J2945     &     \makecell[l]{On-Board System Requirements \\for V2V Safety Communications}    &  Safety\&Assurance     \\ \hline
  12           &   SAE J3048     &     \makecell[l]{Driver-Vehicle Interface Consi-\\derations for Lane Keeping \\Assistance Systems}    &   Safety\&Assurance    \\ \hline
  13           &   SAE J3061     &     \makecell[l]{Cybersecurity Guidebook for \\Cyber-Physical Vehicle Systems}    &   Cybersecurity    \\ \hline
  14          &   UL 4600     &     \makecell[l]{Standard for Safety for the \\Evaluation of Autonomous Products}    &    Safety\&Assurance   \\ \hline
\end{tabular*}
\caption{List of standards selected for evaluation.}
\label{table:description}
\end{table}

\begin{table}[htp]
\centering
\begin{tabular*}{\textwidth}{c cccc c }
\hline

\textbf{\hspace{1pt}No.\hspace{5pt}} & \makecell[c]{\textbf{Time}\\\textbf{dynamicity}}\hspace{10pt} & \makecell[c]{\textbf{Context}\\\textbf{dynamicity}}\hspace{10pt}  & \makecell[c]{\textbf{Collaboration}\\\textbf{dynamicity}}\hspace{10pt} & \makecell[c]{\textbf{Tools}\\\textbf{dynamicity}} \hspace{10pt} & \makecell[c]{\textbf{Ethics}}\\

\hline
\hline

1 &   \multicolumn{1}{c }{-} & \multicolumn{1}{c }{-}   & \multicolumn{1}{c }{-}  & \checkmark &   -  \\ \hline
2 &   \multicolumn{1}{c }{-} & \multicolumn{1}{c }{ \checkmark }   & \multicolumn{1}{c }{-}  & - & - \\ \hline
3 &   \multicolumn{1}{c }{ \checkmark } & \multicolumn{1}{c }{ \checkmark }   & \multicolumn{1}{c }{ \checkmark }  & \checkmark &  \checkmark \\ \hline
4 &   \multicolumn{1}{c }{-} & \multicolumn{1}{c }{ \checkmark }   & \multicolumn{1}{c }{-}  & - & -  \\ \hline
5 &  \multicolumn{1}{c }{-} & \multicolumn{1}{c }{ \checkmark }   & \multicolumn{1}{c }{-}  & - & -  \\ \hline
6 &   \multicolumn{1}{c }{ \checkmark } & \multicolumn{1}{c }{ \checkmark }   & \multicolumn{1}{c }{ \checkmark }  & \checkmark &  - \\ \hline
7 &   \multicolumn{1}{c }{ \checkmark } & \multicolumn{1}{c }{ \checkmark }   & \multicolumn{1}{c }{ \checkmark }  & \checkmark &  - \\ \hline
8 &  \multicolumn{1}{c }{-} & \multicolumn{1}{c }{ \checkmark }   & \multicolumn{1}{c }{-}  & - & \checkmark  \\ \hline
9 &   \multicolumn{1}{c }{ \checkmark } & \multicolumn{1}{c }{ \checkmark }   & \makecell[c]{ unclear collaboration \\form and purpose }  & \checkmark &  \checkmark \\ \hline
10 &   \multicolumn{1}{c }{-} & \multicolumn{1}{c }{-}   & \makecell[c]{ unclear collaboration \\form and purpose }  & - & -  \\ \hline
11  &   \multicolumn{1}{c }{\checkmark} & \multicolumn{1}{c }{-}   & \makecell[c]{ unclear collaboration \\form and purpose }  & \checkmark & \checkmark \\ \hline
12  &   \multicolumn{1}{c }{-} & \multicolumn{1}{c }{-}   & \makecell[c]{ unclear collaboration \\form and purpose }  & - & - \\ \hline
13 &   \multicolumn{1}{c }{ \checkmark } & \multicolumn{1}{c }{ \checkmark }   & \multicolumn{1}{c }{-}  & \checkmark & -  \\ \hline
14 &   \multicolumn{1}{c }{ \checkmark } & \multicolumn{1}{c }{ \checkmark }   & \makecell[c]{ unclear collaboration \\form and purpose }  & \checkmark &  \checkmark \\ \hline
\end{tabular*}
\caption{Evaluation of certification standards. The numbers of standards match with the list of standards defined in Table~\ref{table:description}.}
\label{table:evaluation}
\end{table}

\subsection{Time dynamicity}
Of the documents that were analyzed, only seven demonstrate readiness for the dynamicity in the time context. This means that these standards take further upgrades to a system that has already been certified into account and do not demand that the authority complete the entire certification procedure again. These standards often include instructions for producing the required change paperwork, conducting exhaustive testing, and setting up on-road monitoring to ensure that the update has not interfered with the vehicle's intended operation (UL 4600). 

The other standards (such ISO 15037 and ISO 22735) do not include any information regarding software upgrades at all. Due to this, it is unclear how to proceed in the event that software modifications are required, such as assuring the identification of a new type of traffic sign, and whether the granted certificate will remain valid following an update to the software that has already received certification.

\subsection{Context dynamicity}
Dynamism within the context category, in contrast to the others, appears to be very adequately addressed. Most standards for system verification call for performing simulation tests with unexpected and unpredictable outcomes to fully evaluate the safety or security of the system throughout the design phase of an autonomous vehicle system. Some standards, like ISO 21448, include comprehensive instructions on how to validate the system's performance under ambiguous circumstances. In the papers under evaluation, public road testing is also frequently utilized as a testing technique in the last phases of the development process. However, a worldwide legislative framework for public road AV testing yet to be developed~\cite{bakar2022} as methods and specifications for public road-testing regulations vary between countries. 

Standards that do not support context dynamicity and do not have a check mark in the table typically rely only on a predefined set of tests to verify a system's functionality (ISO 11270, SAE J3018), or they do not include information about testing tools due to the different standard's focus area (ISO/TS 5255).

\subsection{Collaboration dynamicity}
In terms of the collaboration dynamicity dimension, standards ISO/DIS 26262 and ISO 21448 offer a thorough specification and design of communication between the vehicle and other entities within the surrounding ecosystem. Additionally, the ISO/SAE 21434 standard includes dispersed cybersecurity operations that divide cybersecurity duties among several parties. These three standards were designated as completely supportive of dynamic collaboration because they expressly address the potential for collaboration of multiple agents within the ecosystem. 

Although these publications address the dynamic formation of coalitions, there is certainly need for improvement. We note that none of the standards under consideration take any kind of trust management into account while working with other entities. A substantial security risk to the operation of the entire ecosystem is posed by blindly trusting any entity prepared to cooperate since they may have evil motives. 

Terms like \textit{"dependencies between items"}, \textit{"vehicle to vehicle communication"}, or \textit{"arrays of systems implementing other vehicle level functions"} have been discovered in the UL 4600, ISO/TR 4804, ISO/TS 5255, SAE J2945, and SAE J3048 standards. Thus, at least a portion of the issue of vehicle communication with other entities in their surroundings is addressed. However, it is unclear (1) how and in which directions the information is conducted, and (2) if it is feasible to develop a common plan. Because of that, these standards were labeled as offering only limited support for the collaboration dynamicity dimension. All other standards do not seem to directly support collaboration dynamicity at all, or just to a negligible extent.

\subsection{Tools dynamicity}
In more than half of the assessed standards, it is possible to use dynamic tools to determine if a system complies with a certain standard. Such a compliance check often entails on-site testing or the implementation of additional system verification and validation procedures. A significant portion of compliance checks are still carried out by static tools, such as manual examination of documents, processes, and procedures, even when dynamic tools are provided. Standards that are not included in the assessment table as being supportive of the use of dynamic tools utilize only static methods to verify their compliance.

\subsection{Ethics}
The ISO/TR 21959 standard series is largely focused on ethics. Aside from ethical issues, the guidelines also outline the best practices for building social trust. They specifically set forth suggestions for greater adoption of autonomous cars based on an examination of numerous human aspects that affect how people perceive and develop confidence in autonomous systems in the automobile industry.

However, ethical considerations do not appear to be generally taken into account by the reviewed standards. Four criteria in the Safety\&Assurance domain pertain to the need for the absence of unreasonable risk, which is defined as  \textit{"unacceptable in a certain context according to valid societal moral concepts"}~\cite{iso21448}. Other moral issues do not appear to be addressed, and we find the lack of examination of moral issues particularly troubling. We are persuaded that autonomous systems, such as autonomous automobiles, carry a moral obligation since they have direct accountability for human lives.

The ethics of AI systems in general is a widely debated topic nowadays, and authorities are working on sophisticated market regulatory measures. The European Commission's operations might be used as an example. Beginning with the publication \textit{"Policy and Investment Recommendations for Trustworthy Artificial Intelligence"}~\cite{trustworthy-ai-2019}, the European Commission established the groundwork for AI legislation in the European Union, addressing key legal and ethical issues. Several publications on ethics have been produced since then, but many issues and questions remain unresolved.

Because ethics is a complex issue, it is probable that the standards' authors in the field of autonomous driving are expecting further suggestions from central regulatory agencies and will alter the documents once the direction of the regulation is apparent. In any case, autonomous systems cannot be built unless human and social demands are considered throughout system operation. We are certain that, even on a technological level, systems that can control the compliance of ethical standards and that can be followed up on once additional rules are released by the central authorities are required.

\subsection{Evaluation summary}
The evaluation outcomes are varied. While some standards clearly demonstrate very low preparedness to reflect the dynamicity of future autonomous systems and the ecosystems in which they shall operate (e.g., ISO 22735, ISO 22737, or SAE J3018), others demonstrate reasonable readiness for autonomous driving after meeting vast majority (ISO/TR 4804, UL 4600) or all (ISO 21448) of the specified requirements. Even in this scenario, however, our analysis has shown that the future is about to bring further expectations of these certifications that are currently being neglected.
Specifically, we found the absence of contemplating the establishment of dynamic coalitions as well as the absence of trust management in standards with support for coalition dynamicity as one of the most important weaknesses from the assessed features in all of the examined standards. Aside from that, we surely perceive room for improvement in the dimension of ethics. We see that most standards are narrowly focused, mainly on safety, while disregarding ethical elements, which are often directly related.

\section{Future Vision of Dynamic Certification Schemes}
\label{sec:solution}


Given the identified shortcomings of the certification standards in terms of coping with both the certified entities' and their environment's dynamicity, we present six key suggestions for improving the standards' compatibility with the requirements of future dynamic autonomous systems and ecosystems, including enhanced guarantee of ethical awareness.

\subsection{Real-time validation of the certificate's properties}
Traditional certificates, whose objective is to offer certain assurances regarding the quality of the certified system, are issued at a specified moment in time, often shortly before the system is introduced to the market. However, in a dynamic and ever-changing environment, such a certificate's validity may not be maintained over time; rather, the level and conditions of the certified entity may change based on the context or deteriorate with time.

To address the issue of time dynamicity, certification approaches for dynamic autonomous systems must provide for some sort of detection or measurement of certified entity's properties and their changes. In other words, in a dynamic context, the certified properties must be regularly re-checked, and the extent to which the originally awarded certificate and its assurances can be trusted under such changing conditions must be determined.

Consider the following scenario: an AV has a valid traditional certificate, but the vehicle is moving in an unexpected way, raising the suspicion that it has malicious intentions. Other cars will be able to respond to such a scenario more successfully if methods for analyzing suspicious behaviors that differ from expected, usual behavior are in place. In this circumstance, the real-time check would reveal that the car was not certified to handle this exact condition. Another possible result of the real-time check could be that the vehicle's certificate was issued for a software version that is deprecated. In such a case, the originally certified properties have changed and therefore the vehicle cannot be fully trusted without further assurance.

\subsection{Supporting dynamic trust evaluation}
In its essence, certificates have always been meant as a tool to support decisions on whether an entity can be trusted. However, in its current form, static certificates are no longer a sufficient trust-assurance concept, given the complex and highly dynamic environments the systems are operating in. The ever-changing environments of the future autonomous systems indicate that there is a need for autonomous systems to constantly re-evaluate their level of trust in other entities which they interact with. 

As discussed in the previous paragraphs, the level of trustworthiness cannot be unambiguously determined based solely on the issued certificate. It has to be determined based on entity's dynamic properties at the moment of mutual interaction in order to build trust in those certified properties. Trust governance should therefore focus on establishing mechanisms that allow autonomous systems to assess and verify each other's trustworthiness in real-time, based on the entity's dynamic behavior and current context.

This shift in focus towards dynamic trust evaluation leads to rethinking the original certification schemes. Rather than being solely centered around static certificates, the emphasis shall be on establishing frameworks that enable ongoing trust assessment, so that entities can demonstrate their certified properties and their ability to maintain trustworthiness in various scenarios and interactions. 

\subsection{Certificate combined with vehicle's reputation}

Choosing who to trust in dynamic ecosystems becomes a difficult challenge. As shown by research of trust in other disciplines, the concept of reputation is utilized as one of the techniques for building trust. Reputation can be defined as \textit{the overall quality of an entity derived from the judgements by other entities in the underlying network, which is globally visible to all members of the network}~\cite{trust-and-reputation}. When such information is propagated through the network of connected entities, it can have a significant effect on decision-making. Furthermore, by giving information that distinguishes between trustworthy and untrustworthy nodes, reputation can aid in dealing with observed misbehavior~\cite{rep-def} and reducing harm in the event of an insider assault~\cite{reputation1}. 

This concept may be very effective in tackling the time dynamicity issue in certification. In the same way that humans achieve reputation by having their activities judged by their peers, an entity's reputation in a smart ecosystem is determined by how it behaves and interacts with others~\cite{buhnova2022tutorial}. Furthermore, a collection of experience during runtime feeding updates of the trustworthiness score might be utilized to promote or demote the validity of the certified properties. 

Besides, other entities may be able to respond to changes in a dynamic environment in a more flexible manner by complementing decision-making processes about a system's trustworthiness with its reputation. In a hypothetical situation, a vehicle would be less trustworthy if, despite having valid certified properties, its reputation was declining due to other cars reporting its suspicious behavior (showing a software defect or an attack). In the event of a significant reputation decline, the vehicle could be declared temporarily or permanently untrustworthy to avoid additional harm.

\subsection{Extension of the certificate properties to a scale}
Another suggestion to support better decision-making processes based on trust in dynamic environments is to redefine the concept of certificate properties in terms of the range of possible values it can gain. Nowadays, the typical view of a certificate interprets it as a binary value (\textit{valid/invalid}, or \textit{granted/not granted} certificate). Such a perspective is quite restrictive. 

Therefore, we propose redefining the concept as a scale to better describe the current state of the properties of the system installed in an autonomous system. Indeed, there are several interpretations of this notion. The precise meaning of the certification scale has to be researched further, but to name a few possibilities, the scale may e.g. reflect the number of software updates installed in the system or the period of time since the last formal verification of the system's compliance with a certain standard, or simply reflect the reputation discussed in the previous section. 

\subsection{Considering certificate's context-dependant validity}
Evaluating the appropriate amount of trust, or eventually the vulnerability risks, during runtime is strongly context-dependent. Therefore, even when dealing with the context and collaboration dynamics issue, trust management must be implemented. 

Consider two AVs initiating mutual communication. In scenario A, the AVs want to communicate a batch of weather-related data before closing the communication. In scenario B, the entities establish a link with the goal of forming a vehicle platoon to save fuel owing to decreased air resistance. In Scenario B, however, the close and long-term collaboration required for vehicle platoon formation creates more significant trust problems regarding the safety of riding in such close proximity than the circumstance of exchanging data in Scenario A. Even if a vehicle is approved to handle interactions with other cars appropriately in some situations, its behavior in other contexts may not be tested or guaranteed. As a result, before participating in any form of communication, the system itself must be able to determine if the awarded certificate and the other entity can be trusted in the particular context.

\subsection{Certificates combined with ethical concerns}
Evaluating a product's safety or environmental impact before allowing it to reach the market is an increasingly common procedure. However, an assessment of current guidelines in this paper revealed that ethical considerations are still rarely considered. However, technological advancement must be morally monitored. Otherwise, intelligent technologies designed to aid can readily be used to injure or disadvantage specific groups of people.

To solve the certification gap in autonomous system ethics, we propose assessing it in the same manner as a product's safety or environmental concerns are assessed. We propose, in particular, merging certification with \textit{Ethical Digital Identities} (EDI), a notion introduced in~\cite{edi2022}. EDI, which is derived from the notion of \textit{Digital Identities}~\cite{digital-identities-2005}, serves as the foundation for ethically safeguarding the emergence of intelligent safety-critical systems.

\section{Multi-Layer Trust Governance Framework to Support Dynamic Certification}
\label{sec:arch}
It is evident that the current certification schemes need to evolve in order to better reflect the dynamic nature of the future autonomous ecosystems. As discussed in Section~\ref{sec:solution}, mechanisms such as trust management, real-time monitoring, calculation of reputation scores, or context considerations will need to be secured for the realization of the adjustments in the current certification scheme. Such a change can be, however, hardly achieved without a quality governing model securing and guarding these mechanisms. Therefore, we outline a vision of a conceptual multi-layer trust governance framework, which could serve as a base for the implementation of various dynamic and trust-based processes in autonomous ecosystems, including novel certification schemes.

\begin{figure*}[t]
\centering
\includegraphics[width=330px]{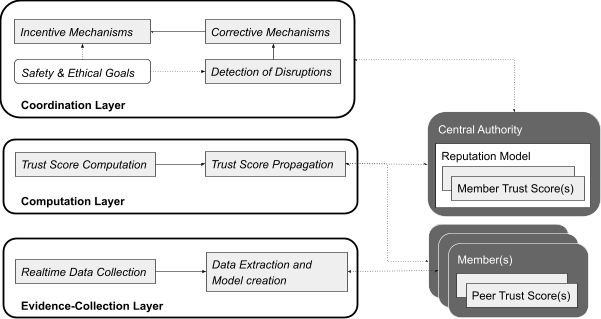}
\caption{Proposed Multi-Layer Trust Governance Framework Architecture}
\label{fig:architecture}
\end{figure*}

In the proposed trust governance framework, we envision a structured architecture organized into three cohesive layers. Each layer contributes vital components that collectively form a comprehensive framework. The envisioned concepts and supporting mechanisms, which are essential for the collection, computation, and propagation of trust within an autonomous ecosystem, are illustrated in Figure~\ref{fig:architecture}.

\subsection{Framework Architecture}

The proposed framework's architecture has a hierarchical structure and consists of the following layers (described from the bottom towards the top):

\paragraph{1. Evidence-Collection Layer} 
The aim of the \textit{Evidence-Collection Layer} is to collect evidence for the current state of the certified properties that promote trust building. In particular, the layer first gathers sufficient trust-related data from real-time interactions among ecosystem members. Second, this data together with the observed ecosystem interactions is used to build desing-time models to support further reasoning about the evidence.

\paragraph{2. Computation Layer} 
The \textit{Computation Layer} focuses on converting the gathered trust-related evidence into individual ecosystem members' trust scores, and managing them. This also includes solving issues such as initialization of an entity's trust score without any previous experience, as well as trust score erosion when the entity does not participate in any interactions for a longer period of time. Besides that, the \textit{Computation Layer} propagates these trust scores throughout the ecosystem to serve multiple purposes, including feeding a reliable, centrally managed reputation model, as well as facilitating the collection of peer opinions among ecosystem members. 

\paragraph{3. Coordination Layer} 
Based on the evidence data and trust scores computed by the previous layers, the \textit{Coordination Layer} focuses on monitoring the situation, checking compliance with the rules, and triggering incentive mechanisms. By implementing the incentive mechanisms, ecosystem members are encouraged to exhibit positive behavior in accordance with the goals set by the ecosystem, such as in terms of safety or ethics. The \textit{Coordination Layer} also addresses the detection and isolation of malicious members who attempt to disrupt the safety of the ecosystem, or disrupt the ecosystem as a whole, and to minimize the impact of their actions.

\bigskip
The proposed framework, which the novel certification schemes could be built upon, offers multiple improvements in terms of ecosystem safety, decision-making processes and ethical compliance. By providing constant monitoring of the situation, evaluating potential threats, and spreading the updated information across the network enables the ecosystem entities to better adapt to ever-changing conditions in real-time. Considering the example of autonomous vehicles, knowing this information is essential for effective and safe handling of various road situations, such as facilitating coordinated maneuvers (forming vehicle platoons) and interactions of the road in general, both in states of emergency (road accident) and normal road conditions (optimizing routes to avoid traffic jams). 

Besides efficiency and safety, the proposed framework impacts the fairness and ethical compliance within the ecosystem, too. Implementing incentives and activating reward or punishment mechanisms in case of (non-)compliance with the ecosystem goals, rules, and ethical principles would support entities to adhere to general moral standards, and to act towards the ecosystems' common good.

\section{Conclusion}
In this paper, we evaluated the readiness of current certification standards for future autonomous driving systems. Taking into consideration the characteristics of both autonomous systems and the dynamic software ecosystems in which they operate, we derived a set of requirements, namely \textit{Time Dynamicity, Context Dynamicity, Collaboration Dynamicity, Tools Dynamicity}, and \textit{Ethics}, and used them to assess the selected certification standards.

Based on the evaluation results, we conclude that the standards are not entirely ready for the expansion of autonomous driving systems, and provide a list of their key shortcomings that we have identified. One of the most serious issues deals with the \textit{Collaboration Dynamicity} aspect, meaning the standards lack to support the creation of dynamic coalitions, as well as the complete absence of any kind of trust management strategies for establishing communication with other entities. Another shortcoming concerns neglecting ethical aspects in the standards' focus.

In order to address the identified shortcomings of the certification standards, we outlined a two-way improvement strategy. In order to help the discussion move towards a comprehensive solution to the identified problems, we first present six suggestions for rethinking certification. By following them, the standardization for autonomous systems in general, not just those in the automotive industry, could better meet the needs of future dynamic systems and ecosystems with ethical awareness, so that these systems can be trusted. Second, we presented an envisioned conceptual Multi-Layer Trust Governance Framework, which aims to establish a governance structure for autonomous ecosystems and related processes, including novel certification schemes. The proposed framework consists of three layers, each serving a specific purpose in ensuring safe and ethical operation of these systems.

\section*{Acknowledgment}

The work was supported by Grant Agency of Masaryk University (GAMU), Interdisciplinary Research Projects sub-programme, project "Forensic Support for Building Trust in Smart Software Ecosystems" (no. MUNI/G/1142/2022).

%
%
%

\bibliographystyle{splncs04}
{\small\bibliography{references}}





\end{document}